\documentclass[twocolumn]{aastex}

\usepackage{graphicx}
\usepackage{enumitem}
\setitemize{noitemsep,topsep=0pt,parsep=0pt,partopsep=0pt}
\usepackage{booktabs}
\usepackage{amsmath}
\usepackage{hyperref}

\shortauthors{Wu et al}

\begin{document}

\title{Evidence for a Nonzero Eccentricity Superpuff Exoplanet WASP-107 b Using JWST Occultation Observation}

\correspondingauthor{Bo Ma}
\email{mabo8@mail.sysu.edu.cn}

\author{Yunke Wu}
\affil{School of Physics and Astronomy, Sun Yat-sen University, Zhuhai 519082, China; {\it mabo8@mail.sysu.edu.cn}}
\affil{CSST Science Center for the Guangdong-Hong Kong-Macau Great Bay Area, Sun Yat-sen University, Zhuhai 519082, China}

\author{Zixin Zhang}
\affil{School of Physics and Astronomy, Sun Yat-sen University, Zhuhai 519082, China; {\it mabo8@mail.sysu.edu.cn}}
\affil{CSST Science Center for the Guangdong-Hong Kong-Macau Great Bay Area, Sun Yat-sen University, Zhuhai 519082, China}

\author{Xinyue Ma}
\affil{School of Physics and Astronomy, Sun Yat-sen University, Zhuhai 519082, China; {\it mabo8@mail.sysu.edu.cn}}
\affil{CSST Science Center for the Guangdong-Hong Kong-Macau Great Bay Area, Sun Yat-sen University, Zhuhai 519082, China}

\author{Zhangliang Chen}
\affil{School of Physics and Astronomy, Sun Yat-sen University, Zhuhai 519082, China; {\it mabo8@mail.sysu.edu.cn}}
\affil{CSST Science Center for the Guangdong-Hong Kong-Macau Great Bay Area, Sun Yat-sen University, Zhuhai 519082, China}

\author{Wenqin Wang}
\affil{School of Physics and Astronomy, Sun Yat-sen University, Zhuhai 519082, China; {\it mabo8@mail.sysu.edu.cn}}
\affil{CSST Science Center for the Guangdong-Hong Kong-Macau Great Bay Area, Sun Yat-sen University, Zhuhai 519082, China}

\author{Shangfei Liu}
\affil{School of Physics and Astronomy, Sun Yat-sen University, Zhuhai 519082, China; {\it mabo8@mail.sysu.edu.cn}}
\affil{CSST Science Center for the Guangdong-Hong Kong-Macau Great Bay Area, Sun Yat-sen University, Zhuhai 519082, China}

\author{Cong Yu}
\affil{School of Physics and Astronomy, Sun Yat-sen University, Zhuhai 519082, China; {\it mabo8@mail.sysu.edu.cn}}
\affil{CSST Science Center for the Guangdong-Hong Kong-Macau Great Bay Area, Sun Yat-sen University, Zhuhai 519082, China}

\author{Dichang Chen}
\affil{School of Physics and Astronomy, Sun Yat-sen University, Zhuhai 519082, China; {\it mabo8@mail.sysu.edu.cn}}
\affil{CSST Science Center for the Guangdong-Hong Kong-Macau Great Bay Area, Sun Yat-sen University, Zhuhai 519082, China}

\author{Bo Ma}
\affil{School of Physics and Astronomy, Sun Yat-sen University, Zhuhai 519082, China; {\it mabo8@mail.sysu.edu.cn}}
\affil{CSST Science Center for the Guangdong-Hong Kong-Macau Great Bay Area, Sun Yat-sen University, Zhuhai 519082, China}

\nocollaboration

\begin{abstract}

WASP-107~b is an extremely low-density super-puff exoplanet whose inflated radius and evidence of strong internal heating make it a key target for understanding planetary structure and evolution.
Its orbital eccentricity is a critical parameter for testing mechanisms such as tidal heating and high-eccentricity migration, yet previous measurements have remained inconclusive.
Due to the large radial velocity jitter caused by stellar activity, and the presence of at least one additional planet in the system, previous radial velocity measurements could not robustly determine the eccentricity of WASP-107~b.
Here we combine the new JWST secondary eclipse data with transit timing data from HST, TESS, and JWST to measure the eccentricity of WASP-107~b. 
Our joint analysis shows that WASP-107~b has an eccentricity of $0.09\pm0.02$, a mass of $0.096\pm0.005 \, M_J$, and an orbital period of $5.721487\pm0.000001$~days. 
We find the $99.7\%$ lower limit of the eccentricity is about 0.04.
These new measurements are consistent with the scenario in which WASP-107~b is in the final stage of high-eccentricity migration. 
Preliminary estimate shows that eccentricity-driven tidal dissipation can provide a significant contribution to the energy required to sustain the observed radius inflation of WASP-107~b.
Our results establish the dynamical status of one of the most intriguing low-density exoplanets known, and offer new insights into its formation and evolution history.
\end{abstract}


\section{Introduction}

Over 6,000 exoplanets have been confirmed to date, predominantly detected through the transit method\footnote{\href{https://exoplanetarchive.ipac.caltech.edu}{https://exoplanetarchive.ipac.caltech.edu}}.
WASP-107~b presents a unique laboratory for exoplanet studies with its Jupiter-like radius and Neptune-like mass, yielding exceptionally low bulk density \citep{anderson_discoveries_2017, dai_oblique_2017}. Its extended atmosphere exhibits significant scale height, enabling detailed atmospheric characterization including detections of helium in the eroding atmosphere \citep{spake_helium_2018}, with subsequent molecular detections of water vapor, sulfur dioxide, silicates, and methane from JWST observations \citep{welbanks_high_2024,dyrek_so2_2024}. 

Despite extensive atmospheric characterization, the formation and evolution of WASP-107~b remain uncertain.
The planet currently occupies a slightly retrograde, nearly polar orbit \citep{rubenzahl_tesskeck_2021,dai_oblique_2017}.
\citet{petrovich_disk-driven_2020} proposed that disk-driven resonance migration could explain the existence of close-in planets on highly inclined orbits, 
while \citet{yu_are_2024} suggested an alternative scenario in which WASP-107~b underwent high-eccentricity migration followed by ongoing orbital decay \citep[see also][]{piaulet_wasp-107bs_2021, Sethi2025, batygin_tides_2025}. 
The planet’s unusually large radius has motivated various explanations, including tidal dissipation within the planet’s interior \citep{sing_warm_2024, Sethi2025, welbanks_high_2024} and Ohmic heating driven by atmospheric dynamics and magnetic interactions \citep{dyrek_so2_2024, batygin_tides_2025}.

A key orbital parameter for constraining the evolutionary state of WASP-107~b is its eccentricity. 
Previous measurements by \citet{piaulet_wasp-107bs_2021} reported an eccentricity of $e = 0.06 \pm 0.04$, leaving open the possibility of a circular orbit, and more recently \citet{Yee2025} presented radial-velocity data consistent with zero eccentricity for WASP-107~b.
This large uncertainty has so far prevented a definitive understanding of the planet’s migration history and internal heating energy sources. Measuring secondary eclipse has traditionally been used to deduce a precise measurement of the orbital eccentricity \citep{Deming2005}. 
It would simultaneously improve our estimate of WASP-107~b’s eccentricity while providing a better constraint on the tidal energy dissipation rate in the interior of WASP-107~b . 
Because tidal heating contributions may be significant for its radius inflation, the observation of secondary eclipse would also provide a unique opportunity to test the tidal heating model.

As part of the \textit{Exoplanet Ephemerides CHange Observations} (ExoEcho) project \citep{wang_long-term_2024, ma_exoplanet_2025}, we present a new joint analysis of high-precision primary transit, secondary eclipse, and radial velocity data to measure the orbital eccentricity of WASP-107~b. Our analysis incorporates newly obtained observations from \textit{HST}, \textit{JWST}, and \textit{TESS}, allowing us to achieve an improved orbital solution. 
We confirm that WASP-107~b has a statistically significant non-zero eccentricity and place an upper limit on its orbital decay rate, providing new insight into its tidal evolution and possible heating mechanisms. 
The more precise eccentricity would also allow for a better measurement of the properties of the second planet in this system, WASP-107~c. 

This paper is organized as follows. 
In Section \ref{sec:data and observation}, we describe the observational data sets and light-curve processing methods. 
Section \ref{sec:transit timing variation analysis} uses the transit timing variation (TTV) analysis to constrain the orbital period and conjunction time. 
In Section \ref{sec:ecc-measurement}, we perform a joint fit of the primary transit, secondary eclipse, and radial velocity data, yielding a precise eccentricity measurement for WASP-107~b. 
Finally, Section \ref{sec:discussion} discusses the implications of our new measurements for the formation and tidal evolution of the WASP-107 system, and Section \ref{sec:conclusion} summarizes our main findings.

\section{Data and Observation \label{sec:data and observation}}

Proposal information for all data used in this work is summarized in Table~\ref{tab:data proposal}. We analyzed JWST data from six observations (five transits, one eclipse) spanning four proposals (1185: PI Thomas Greene; 1201: PI David Lafreniere; 1224: PI Stephan Birkmann; 1280: PI Pierre-Olivier Lagage), HST data from two proposals (14915: PI Laura Kreidberg; 14916: PI Jessica Spake), and new TESS Director's Discretionary Target observations (DDT-085; Program ID 085: PIs Zixin Zhang \& Bo Ma).
These datasets yielded a total of 5 transit timings from JWST, 2 from HST, and 2 from TESS. Our team proposed the TESS observations, which began on April 9, 2025, at 19:05:42 UT with a 20-second cadence. All data are publicly accessible through the MAST archive\footnote{\href{https://mast.stsci.edu}{https://mast.stsci.edu}} \citep{doi:10.17909/t97p46}.

\begin{figure*}[h]
    \centering
    \includegraphics[width=0.95\textwidth]{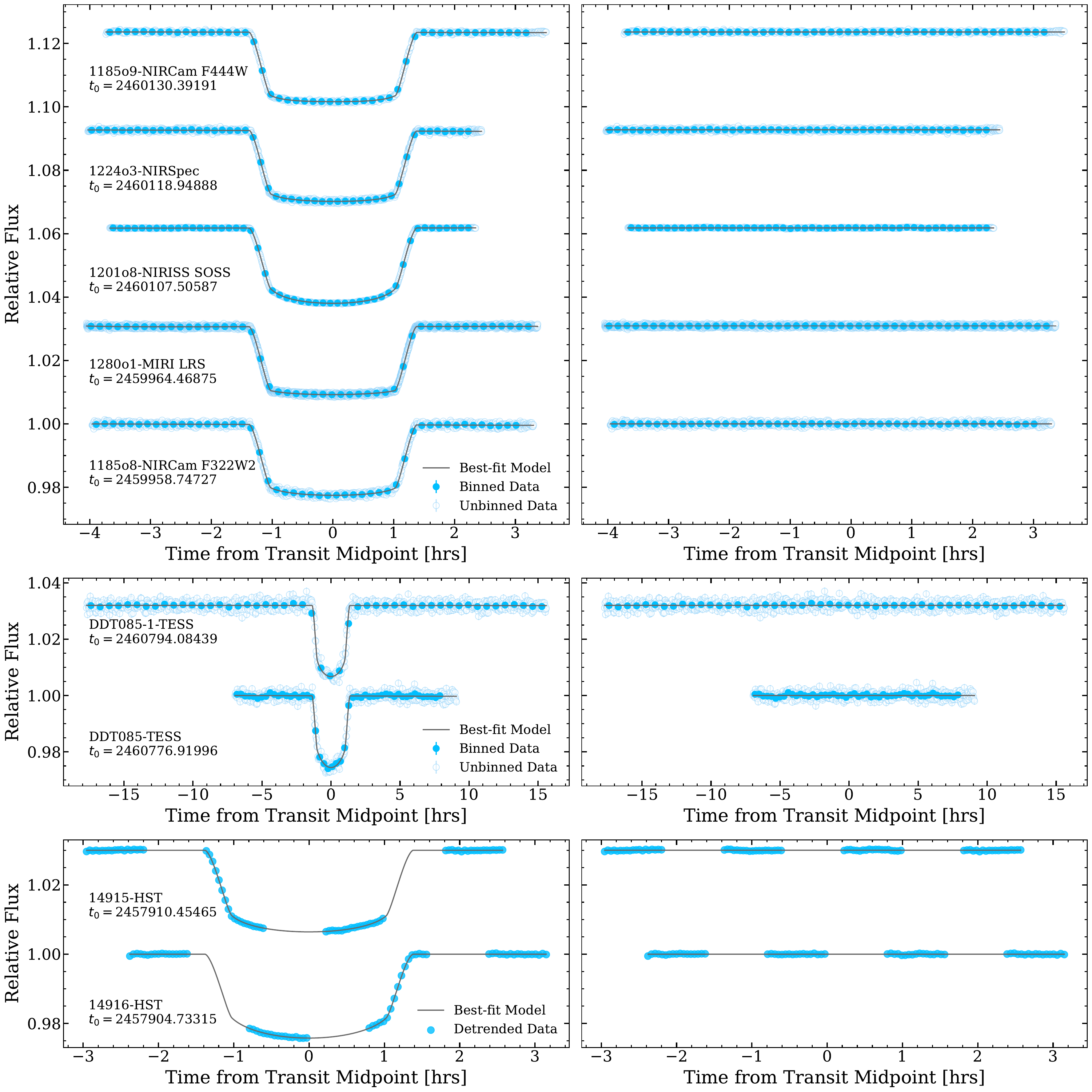}
    \caption{Reduced light curves and transit model fitting results of WASP-107~b. Left panels show relative flux light curves, while right panels display corresponding residuals. Data are presented with vertical offsets for clarity: JWST (top), TESS (middle), and HST (bottom), each labeled by observation ID and instrument. For JWST and TESS, un-filled blue points show unbinned data and solid blue points show 50-point binned data. For HST, solid blue points represent unbinned normalized detrended data. Black lines indicate best-fitting transit models.}
    \label{fig:WASP-107 five white light curves together}
\end{figure*}

We performed light curve analysis separately for each facility, which are described in the following subsections. We presented analysis for the eclipse data from JWST with joint RV data in the next section. 
The reduced light curves and transit model fitting results are shown in Figure \ref{fig:WASP-107 five white light curves together}. The best-fit transit middle time data are listed in Table \ref{tab:WASP-107b full data}.

\begin{deluxetable*}{cccccc}[!h]
\tablecaption{Complete list of transit timings for WASP-107~b \label{tab:WASP-107b full data}}
\tabletypesize{\scriptsize}
\tablewidth{0pt}
\tablehead{
\colhead{$t_{\mathrm{mid}}$ ($\mathrm{BJD_{TDB}}$)} & \colhead{Uncertainty} & \colhead{Source\tablenotemark{a}} & \colhead{Reference} & \colhead{Epoch} & \colhead{Date}
}
\startdata
2457584.329897 & 0.000032 & K2 & \cite{dai_oblique_2017} & 0 & 2016-07-14 \\
2457904.7331517 & 0.0000465 & HST & This work & 56 & 2017-05-31 \\
2457910.4546520 & 0.0000458 & HST & This work & 57 & 2017-06-05 \\
2458579.86858 & 0.0004288 & TESS-I\&W & \cite{ivshina_tess_2022} & 174 & 2019-04-06 \\
2458591.3117 & 0.0003547 & TESS-I\&W & \cite{ivshina_tess_2022} & 176 & 2019-04-17 \\
2459958.7472685 & 0.0000335 & JWST & This work & 415 & 2023-01-14 \\
2459964.4687522 & 0.0000117 & JWST & This work & 416 & 2023-01-19 \\
2460107.5058741 & 0.0000098 & JWST & This work & 441 & 2023-06-12 \\
2460118.9488811 & 0.0000213 & JWST & This work & 443 & 2023-06-23 \\
2460130.3919121 & 0.0000266 & JWST & This work & 445 & 2023-07-04 \\
2460776.9199579 & 0.0002028 & TESS & This work & 558 & 2025-04-11 \\
2460794.0843876 & 0.0002076 & TESS & This work & 561 & 2025-04-28 \\
\enddata
\end{deluxetable*}

\subsection{JWST}

We extracted white light curves from raw JWST data using two pipelines: \texttt{Eureka!} \citep{bell_eureka_2022} and \texttt{exoTEDRF} \citep{radica_exotedrf_2024}. \texttt{Eureka!} is a well-established spectral reduction tool applied to systems including WASP-39~b and TRAPPIST-1 \citep{ahrer_early_2023,rustamkulov_early_2023,alderson_early_2023,howard_characterizing_2023,powell_sulfur_2024}. \texttt{exoTEDRF} (previously \texttt{supreme-SPOON}) has been similarly validated \citep{radica_awesome_2023,feinstein_early_2023,howard_characterizing_2023,lim_atmospheric_2023}. Both pipelines follow similar reduction stages: data preprocessing (flat-field correction, background subtraction, unit conversion, etc.), spectral extraction, light curve fitting, and transmission spectrum derivation. We used both solely for white light curve extraction and transit timing determination, with pipeline settings and light curve results provided online (cf. Section \ref{sec:data availability}). 

We modeled these transits using the \texttt{batman} package \citep{kreidberg_batman_2015}, with priors detailed in Table \ref{tab:LC priors} from previous studies \citep{mocnik_starspots_2017, kokori_exoclock_2023, piaulet_wasp-107bs_2021}. Key parameter constraints were as follows: a uniform prior $\mathcal{U}$(0, 0.3) for $R_p/R_{\star}$; a normal prior for $t_{mid}$ centered on preliminary fits with a standard deviation of 0.01 days; a fixed orbital period $P$; normal priors for $a/R_{\star}$ (with standard deviation 1) and inclination $i$ (with standard deviation 1$^{\circ}$); and fixed eccentricity $e$ at zero. 
For limb darkening, we adopted reparameterized quadratic coefficients $q_1$ and $q_2$, each uniformly constrained between 0 and 1 \citep{kipping_efficient_2013}. 
Our systematics model for each primary transit light curve include a linear trend in time, a linear trend with spatial position and PSF width, and a flux uncertainty multiplier to account for residual noise \citep{bell_nightside_2024,valentine_jwst-tst_2024}. 
We have done tests to show that these systematics correction are important for obtaining the radius ratio between the star and the planet, but has negligible impact on the eccentricity measurement in this study. 
Since the secondary eclipse is expected to be shallow, we only de-trend the light curve around the expected secondary eclipse using a linear polynomial with time. We choose not use a linear de-trending correction using spatial position and PSF width to prevent from over-fitting.
We have conducted timing-averaging tests \citep{pont_effect_2006} to demonstrate the goodness of fit for each transit light curve from the JWST data (Figure \ref{fig:WASP-107b residual rms JWST}).

\subsection{HST}

We obtained HST/WFC3 slitless spectroscopy data (grisms G102 and G141; proposals 14915 and 14916) and reduced them following \cite{ma_exoplanet_2025} using the \texttt{Iraclis} pipeline \citep{Tsiaras2016, Tsiaras2016_55, Tsiaras2018}. Transit times were derived by fitting white light curves with the \citet{Mandel2002} model. All parameters except the mid-transit time ($t_{\mathrm{mid}}$) and planet-to-star radius ratio ($R_p/R_\star$) were fixed to literature values, and limb darkening was modeled nonlinearly \citep{claret_2000} using coefficients from ATLAS models \citep{Howarth2011}.

\subsection{TESS}

We obtained 20-second cadence observations of WASP-107~b through the TESS Director's Discretionary Time program (DDT-085). The  light curve, processed by the \texttt{SPOC} pipeline \citep{jenkins_tess_2016}, contains two transits as shown in Figure \ref{fig:WASP-107b full white light curve TESS}. The data were segmented by transit event, and transit times were derived using the same method and priors as in the JWST analysis. Parameter values are specified in Table \ref{tab:LC priors}.
The time-averaging tests for the TESS data are presented in Figure \ref{fig:WASP-107b residual rms TESS}, to demonstrate
the goodness of transit light curve fitting.

\subsection{Radial Velocity Measurements}
We adopt the RV measurements from Keck/HIRES observations, which are reduced by \citet{piaulet_wasp-107bs_2021}. There are a total of 60 RV measurements (see also \citet{howard_planet_2025}). Since the star is an active K-type dwarf, the measurement error is larger than the photon-limited error. We thus added a quadratic error term $\sigma_H$ during our joint fitting process following the results of \citet{piaulet_wasp-107bs_2021}.
There are another 31 RV data points from CORALIE \citep{anderson_discoveries_2017, piaulet_wasp-107bs_2021}, which have a substantially larger uncertainty of $\sim 10$~m/s. 
For this reason, our fiducial analysis includes only the HIRES RVs in Section~\ref{sec:ecc-measurement}, while results incorporating the CORALIE RV data are presented in Section~\ref{app:coralie} of the Appendix.

\section{Orbital Period and Conjunction Time Measurements \label{sec:transit timing variation analysis}}

Before we run a joint-fitting of the photometry and RV data, we first pin down the orbital period and inferior conjunction time ($T_c$) using the primary transit data of HST, Kepler, TESS, and JWST described in Section \ref{sec:data and observation}. 

\subsection{Method}
To obtain a precise measurement of the orbital period and inferior conjunction time of WASP-107~b, we performed a transit timing variation (TTV) analysis in this section.
We combined our measured transit timings with literature values (listed in Table \ref{tab:WASP-107b full data}) and applied two established models for TTV fitting: a linear model and an orbital decay model. Both approaches have been extensively used in previous studies \citep{maciejewski_departure_2016,patra_apparently_2017,yee_orbit_2019,maciejewski_revisiting_2021,yang_tentative_2022,yang_transit_2024, wang_long-term_2024,ma_exoplanet_2025}.

The linear model assumes a constant orbital period:
\begin{equation}
    t_{\mathrm{mid}}(E) = t_0 + E\times P,
\end{equation}
where $t_{\mathrm{mid}}$ is the mid-transit time, $P$ is the orbital period, $E$ is the epoch number (orbits since reference), and $t_0$ is the reference mid-transit time.

To minimize errors over large epoch ranges, we use the cumulative form:
\begin{equation}
    \label{equation:TTV linear cumulative model}
    t_{\mathrm{mid}}(E) = t_{\mathrm{mid}}(E-1) + P_{E-1}.
\end{equation}

The orbital decay model incorporates a uniformly changing period:
\begin{equation}
    t_{\mathrm{mid}}(E) = t_0 + EP + \frac{1}{2}\frac{dP}{dE}E^2.
\end{equation}

Its cumulative formulation is:
\begin{equation}
    \label{equation:TTV orbital decay cumulative model}
    \begin{aligned}
    t_{\mathrm{mid}}(E) &= t_{\mathrm{mid}}(E-1) + P_{E-1},  \\
    P_E &= P_{E-1} + \dot{P} \left( \frac{P_{E-1} + P_E}{2} \right), \\
    \end{aligned}
\end{equation}
where $\dot{P} = dP/dt = P^{-1} dP/dE$ is the orbital period decay rate.

We evaluated model performance using the Bayesian Information Criterion (BIC) \citep{schwarz_estimating_1978}:
\begin{equation}
\mathrm{BIC} = \chi^2 + k\ln(n),
\end{equation}
where $k$ is the number of parameters (2 for linear, 3 for decay model) and $n$ is the number of data points. A $\Delta\mathrm{BIC} > 10$ between models constitutes strong evidence for preferring the lower-BIC model.
Bayesian inference was performed using \texttt{dynesty} \citep{speagle_dynesty_2020,koposov_joshspeagledynesty_2022} with Dynamic Nested Sampling \citep{higson_dynamic_2019}.

\begin{figure*}
    \centering
    \includegraphics[width=0.85\textwidth]{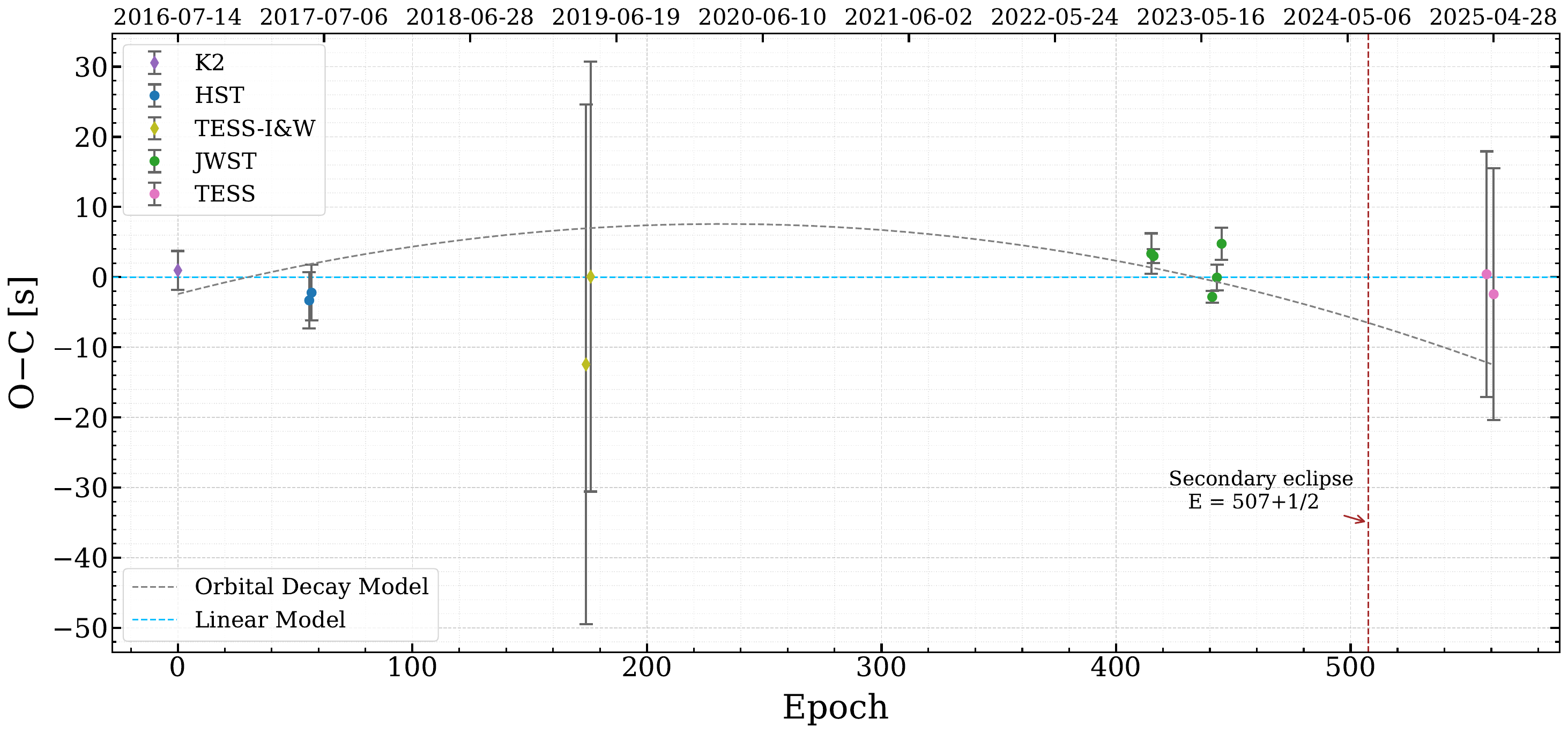}
    \caption{Transit timing variations of WASP-107~b with best-fit linear model and orbital decay model. Blue dashed line shows best-fitting linear model, while gray dashed line indicates best-fitting orbital decay model. Points are color-coded by data source: circular markers denote new measurements from this work (HST, JWST, TESS), and diamond markers represent literature data. The corresponding epoch for the only secondary eclipse observation is shown by a vertical red dashed line.}
    \label{fig:WASP-107b TTV Orbital Decay}
\end{figure*}

\subsection{Modeling Results}

Figure \ref{fig:WASP-107b TTV Orbital Decay} displays the best-fit results for both of the linear model (blue dashed line) performs and the orbital decay model (gray dashed curve).
We find that the linear model can best explain the transit middle time data, which gives us an orbital period of $P_b = 5.7214876\pm0.0000001$~days, and inferior conjuction time $T_{0, b}=2459958.74727\pm0.00003$~BJD.
The orbital decay model gives a substantial orbital decay rate of $-24\pm11$~~ms~yr$^{-1}$ and a small $\Delta\mathrm{BIC} = 2.3$ value, which suggests the linear model performs equally well as the orbital decay model. 
Thus, we decide to keep using the $P_b$ and $T_{0, b}$ from the linear model in our joint-fitting analysis in  Section \ref{sec:ecc-measurement}.

\section{Eccentricity Measurement \label{sec:ecc-measurement}}

\begin{figure}
    \centering
    \includegraphics[width=0.5\textwidth]{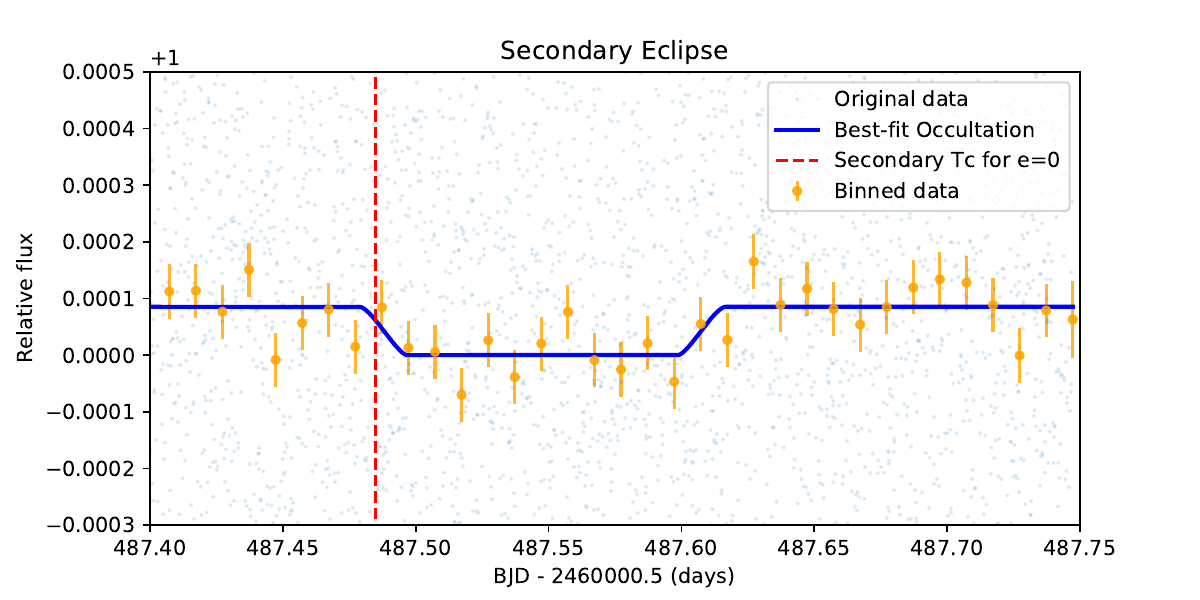}
    \caption{Joint fitting results: Secondary eclipse light curve of WASP-107~b. Detrended JWST measurements are shown as pale blue dots, and binned into the yellow points. The blue curve shows the best-fit occultation (secondary eclipse) model. The red dashed line marks the occultation middle time (superior conjunction) if WASP-107~b has a zero eccentricity. The time shift of the eclipse is very sensitive to the eccentricity, with $\delta t = 2 e_b P_b \cos \omega_b / \pi$. }
    \label{fig:second}
\end{figure}

\begin{figure}
    \centering
    \includegraphics[width=0.45\textwidth]{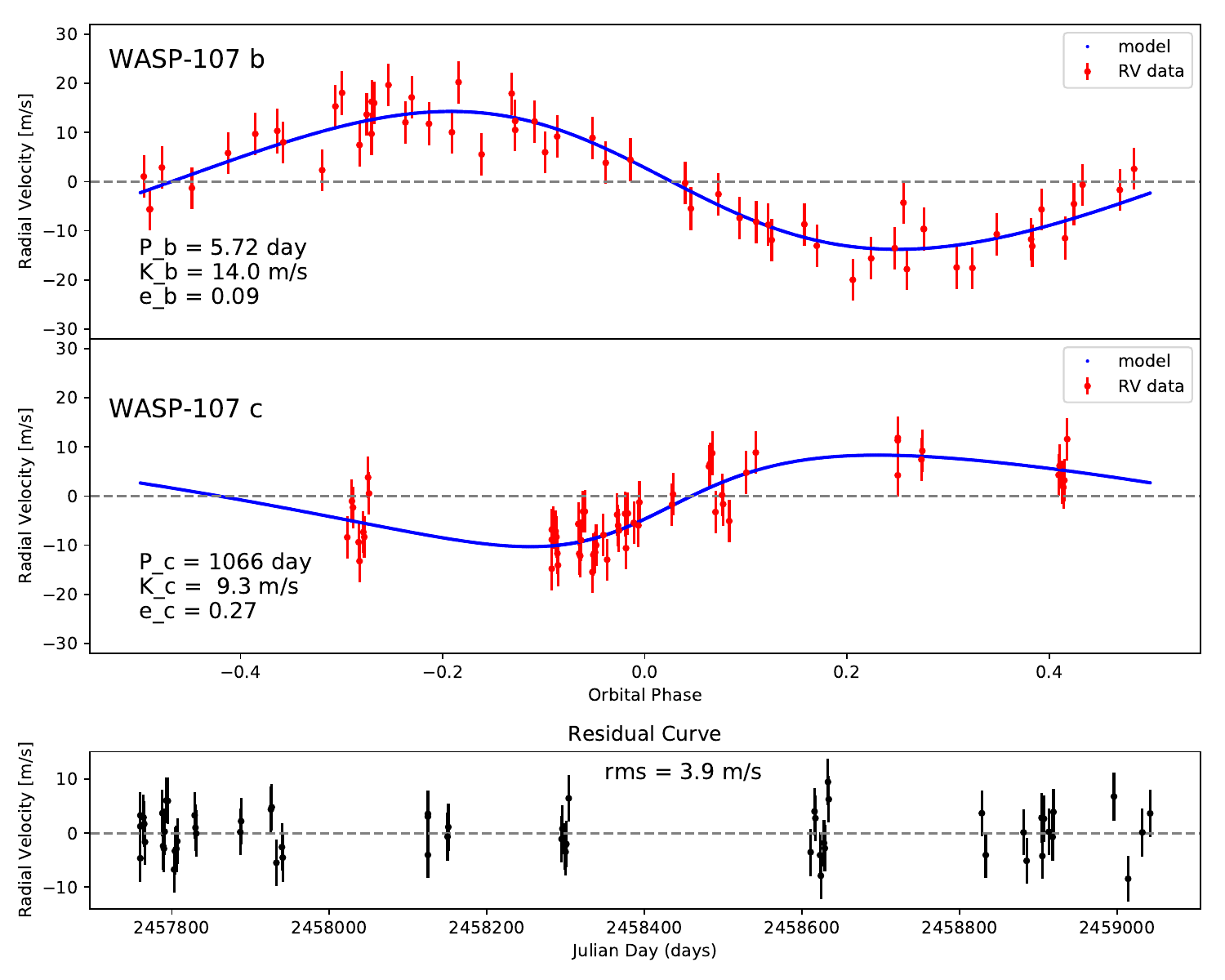}
    \caption{Joint fitting results: RV curves. Maximum-likelihood two-planet Keplerian orbital model for WASP-107. In the top and middle panel, the Keplerian orbital model for the other planet has been subtracted.}
    \label{fig:rv_combine}
\end{figure}

We model the primary JWST transit events, the JWST secondary eclipse event, and the HIRES RV data simultaneously, during which the $P_b$ and $T_{b, 0}$ values are fixed. 
We use $batman$ \citep{kreidberg_batman_2015} package to model the transit and eclipse when fitting the light curves.
The orbital period and inferior conjunction time are taken from Section~\ref{sec:transit timing variation analysis}. Initial guesses of the other parameters ($P_c$, $K_b$, $K_c$, $e_b$, $e_c$, $\omega_b$, $\omega_c$, $R_p/R_\star$, $i_b$, $a/R_\star$) are obtained from from \citet{piaulet_wasp-107bs_2021}, \citet{Knudstrup2024}, and a uniform prior is often adopted. 
We use the \textit{emcee} package \citep{Foreman-Mackey2013} to do the fitting. We utilize 100 walkers for 10,000 steps, where the first 2,000 steps are discarded as burn-in after visually check. 
To assess the convergence of our MCMC sampling, we computed the Gelman-Rubin statistic (\(\hat{R}\)) for all fitted parameters. We obtained \(\hat{R} = 1.03\), which is close to unity and indicates that the chains have mixed well and reached satisfactory convergence. This value of \(\hat{R}\) suggests that the between-chain variance is only slightly larger than the within-chain variance, implying that the posterior estimates derived from our sampler are statistically robust.  
To reduce the run time of the joint fitting, we only use the 1185o8-NIRCam primary transit data in our joint fitting because it shows one of the lowest relative residual RMS among all five JWST transit light curves. 
Our test runs show that switching to the other four transit light curves has negligible effect on the fitting results. 
We show the final best-fitted secondary eclipse data in Figure~\ref{fig:second}, and RV curves in Figure~\ref{fig:rv_combine}. 
We fit the linear systematics model ($c_0+c_1*t$) for the JWST secondary eclipse data simultaneously with the joint fitting.
The best-fitting parameters are summarized in Table~\ref{tab:wasp107_mcmc_params}, and the corner plots are shown in Figure~\ref{fig:HIRES_full_corner}. 
We adopt the host star mass estimate from \citet{piaulet_wasp-107bs_2021} when deriving the planetary masses in Table~\ref{tab:wasp107_mcmc_params}.

\begin{figure}
    \centering
    \includegraphics[width=0.45\textwidth]{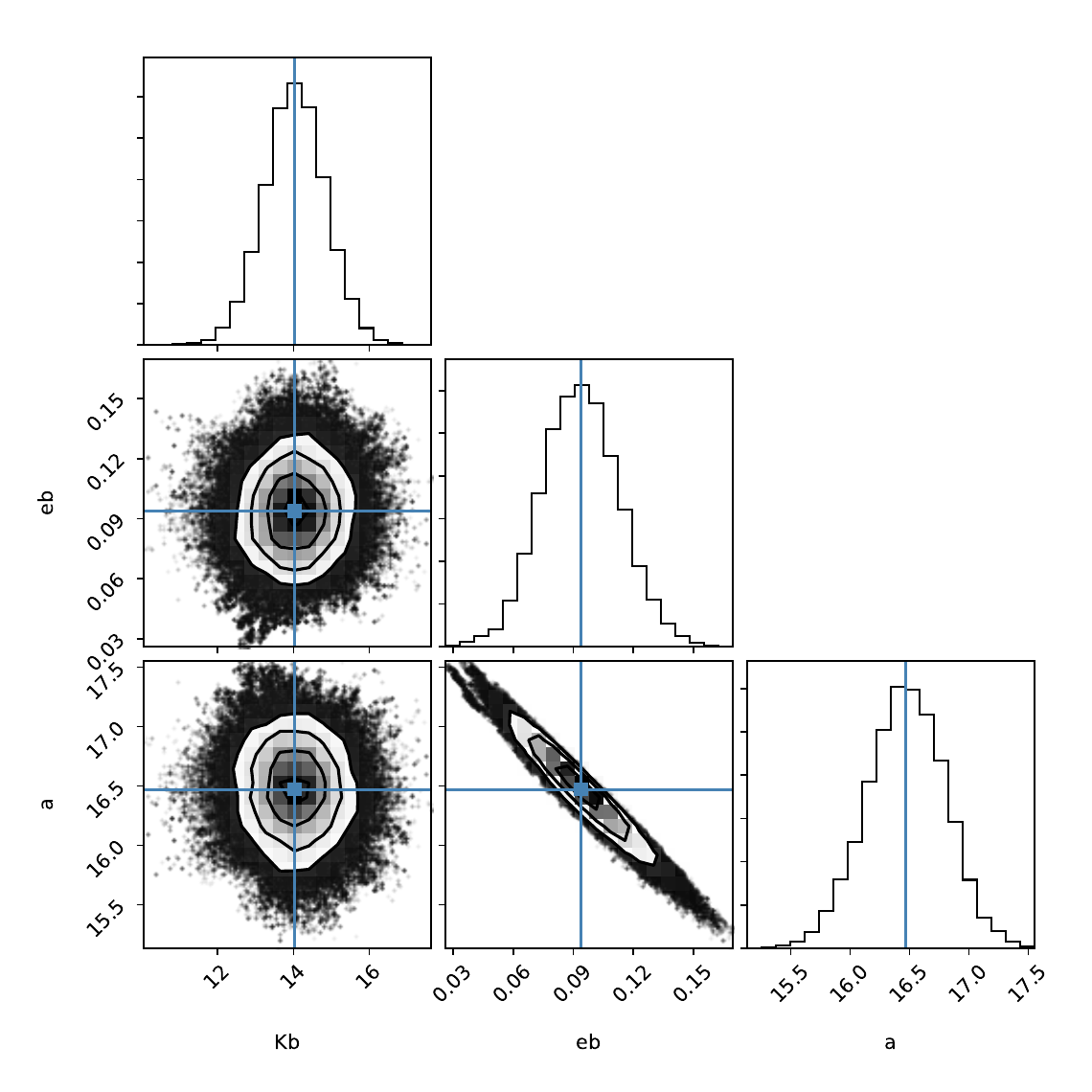}
    \caption{Corner plot showing the correlation between $e_b$ and $a/R_\star$.}
    \label{fig:corner_plot1}
\end{figure}

\begin{deluxetable*}{lcc}[!h]
\tablecaption{MCMC joint-fit results for WASP-107 \label{tab:wasp107_mcmc_params}}
\tablewidth{0pt}
\tablehead{
\colhead{Parameter} & \colhead{Value} & \colhead{Error $(+\sigma\ / -\sigma)$ }
}
\startdata
$P_b$ (day) & $\equiv 5.7214876$ & $\pm0.0000001$ \\
$T_{0,b}$ (BJD-2450000) & $\equiv 9958.74727$ & $\pm3\times10^{-5}$ \\
$a/R_\star$  &  $16.5$  & $\pm0.04$ \\
$K_b$ (m s$^{-1}$) & 14.0 & $\pm0.8$ \\
$e_b$ & 0.09 & $\pm0.02$  \\
$e_b$ & 0.04  & 99.7\% lower limit \\
$e_b$ & 0.15  & 99.7\% upper limit \\
$\omega_b$ (deg) & 79.3 & +2.5 / -2.7 \\
$i_b$ (deg) & 89.55 & +0.25 / -0.18 \\
$M_b$ ($M_{\rm Jup}$) & 0.096 &$\pm0.005$ \\
$R_p/R_*$ & 0.1437 & $\pm0.0003$ \\
$F_p/F_*$ ($10^{-5}$) & 8.5 & $\pm1.8$ \\
$T_{0,c}$ (BJD-2450000) & 8924 & +67 / -71 \\
$P_c$ (days) & 1066 & +24 / -25 \\
$K_c$ (m s$^{-1}$) & 9.3 & $\pm1.0$ \\
$e_c$ & 0.27 &  $\pm0.08$ \\
$\omega_c$ (deg) & 246 & +27 / -30 \\
$M_c \sin i$ ($M_{\rm Jup}$) & 0.35 & $\pm0.04$ \\
$u_1$ & 0.15 & $\pm0.03$ \\
$u_2$ & 0.11 & $\pm0.06$ \\
$\gamma_H$ (m s$^{-1}$) & 1.3 & $\pm0.7$ \\
$\sigma_H$ (m s$^{-1}$) & 4.0 & +0.5 / -0.4 \\
$c_0$ (ppm) & -53 & $\pm14$  \\
$c_1$ (ppm/day) & -1.43E-3 & $\pm9$E-5 \\
\enddata
\tablecomments{
    Parameter definitions: 
    $K_b$/$K_c$ = Radial velocity semi-amplitude of component b/c; 
    $e_b$/$e_c$ = Orbital eccentricity of component b/c; 
    $\omega_b$/$\omega_c$ = Argument of periastron of component b/c; 
    $T_{0,b}$/$T_{0,c}$ = Transit/mid-transit time of component b/c (relative to BJD-2450000); 
    $P_c$ = Orbital period of component c; 
    $\gamma_H$ = Systemic radial velocity; $\sigma_H$ is the HIRES RV jitter term;
    $R_p/R_*$ = Planet-to-star radius ratio; 
    $u_1$/$u_2$ = Quadratic limb-darkening coefficients; 
    $F_p/F_*$ = Planet-to-star flux ratio; 
    $i$ = Orbital inclination; $c_0$ and $c_1$ are the linear systematic parameters for JWST occultation data.
}
\end{deluxetable*}

As seen from Table~\ref{tab:wasp107_mcmc_params}, the eccentricity of WASP-107~b is 0.09$\pm$0.02. This large eccentricity is supported by an offset between the observed middle time of secondary eclipse and the predicted secondary eclipse middle time for a circular orbit ($t_{0, b}+P_b/2$, as shown in Figure~\ref{fig:second}). This offset is proportional to $e_b \cos \omega_b$ \citep{Kopal1946, Alonso18}. The duration of the secondary eclipse is proportional to $e_b \sin \omega_b$. These two constraints from the secondary eclipse data, together with RV data, support an non-zero eccentricity for WASP-107~b. 
We do not apply a light-travel–time correction when modeling the secondary-eclipse timing, as the expected effect is only at the level of ~1~min.
During the fitting, we found a tight correlation between the best-fit $e_b$ and $a/R_{\star}$, which is shown in Figure~\ref{fig:corner_plot1}. 
An increase of 0.2 in $a/R_{\star}$ will result in a decrease of about 0.01 in the best-fit eccentricity of WASP-107~b.
Our joint MCMC analysis yields a 99.7\% lower limit of $e_b > 0.04$ and an upper limit of $a/R_\star < 17.4$ for WASP-107~b.
This is similar to the finding of \citep{Mahajan24} that, strong constraint can be put on the stellar mass and radius when coupling RV, primary transit and secondary occultation data. 

Apart from the linear detrending, decorrelation against the position and full width half maximum (FWHM) of the point spread function (PSF) are often conducted when reducing the secondary eclipse data from JWST. However, to avoid over-correcting the data, we choose not to apply these corrections, which were estimated to be at the level of $\sim$10~ppm. Given the shallow secondary-eclipse signal in the JWST data, we have performed three additional tests to assess its significance. First, we repeat the joint fit by assuming a fixed zero eccentricity. The resulting $\Delta \mathrm{BIC} = 10.5$ relative to our best-fit model, which favors a non-zero eccentricity. Second, we fit a straight line to the JWST data near the expected secondary eclipse by fixing $F_p/F_\star ==0$, instead of using the \textit{batman} eclipse model. This new fit yields $\Delta \mathrm{BIC} = 12.9$ compared to the best-fit model, which supports the detection of a secondary eclipse signal. Third, we searched for additional eclipse-like dips with a width of 3~hr and depth of 85~ppm in the $\sim 40$~hr JWST residual light curves after removing the primary transit and detected secondary eclipse; none were found. This implies that the probability of the detected secondary eclipse arising from pure instrumental systematics is likely to be less than $3/40 \approx 7.5\%$.

We next compare our results with these from \citet{piaulet_wasp-107bs_2021} and \citet{Yee2025}. 
\citet{piaulet_wasp-107bs_2021} have mainly used radial velocity data to obtain an eccentricity of $0.06\pm0.04$ for WASP-107~b, while in our study, the constraints come from a combination of transit, eclipse, and RV measurements. 
Thus, we now have a more precise measurement of the eccentricity of WASP-107~b, and can rule out the zero-eccentricity scenario for WASP-107~b at a $5-\sigma$ significance.

In a recent study, \citet{Yee2025} presented additional radial-velocity measurements and found that WASP-107 b is consistent with zero eccentricity, placing a 95\% confidence upper limit of $e_b < 0.05$. While our measurements place a 95\% confidence lower limit of $e_b > 0.06$. Several factors may contribute to the tension between their results and ours. First, unknown stellar RV jitter may bias the eccentricity measurement of WASP-107~b. Second, the poorly constrained orbital parameters of the outer planet WASP-107~c can introduce additional uncertainty on the measurements of planet WASP-107~b's eccentricity. 
Third, adopting a larger $a/R_\star$ prior value can partially reconcile the discrepancy, as indicated by the correlation between $e_b$ and $a/R_\star$ in Figure~\ref{fig:corner_plot1}. 
For MCMC method, it is well-known that when the observation data are highly constraining, the prior has little influence on the measured value. However, when the observation data are not highly-constraining, the prior can have a large impact on the measured values.
Thus, to improve the eccentricity constraint on WASP-107~b in the future will likely require (1) a more tight $a/R_\star$ prior from asteroseismology, (2) additional secondary-eclipse observations with JWST, and (3) tighter constraints on the orbital parameters of the outer planet WASP-107~c.

\section{Discussion\label{sec:discussion}}
\subsection{Orbital Circularization}

The confirmation of non-zero eccentricity ($e_b = 0.09 \pm 0.02$) of WASP-107~b provides new insights into the ongoing tidal evolution of this dynamically active system. 
For a close-in planet with a period of only 5.7~days, such an eccentricity is expected to be rapidly damped by tidal dissipation within the planet unless it is actively maintained. 
In the study of \citet{piaulet_wasp-107bs_2021}, they have shown that the timescale of tidal circularization for WASP-107~b is about 66~Myrs, much shorter than the estimated age of the system. 
Any remaining eccentricity of the inner planet resulting from dynamical interactions will therefore be quickly dampened by tidal interactions with its host star. 
The persistence of $e_b > 0$ therefore implies either a relatively weak tidal quality factor or recent dynamical interactions that excited the eccentricity to high values, for instance from the outer companion WASP-107~c. 
In the study of \citet{yu_are_2024}, they have proposed that WASP-107~b could be in the final stage of high-e migration and orbital circularization. Our measurement of an eccentricity of $0.09\pm0.02$ is in agreement with their study. 

\subsection{Tidal Inflation}
The large radius of WASP-107~b was initially attributed to a high H/He envelope mass fraction (see, e.g., \citet{piaulet_wasp-107bs_2021}). However, recent independent analyses of JWST spectra by \citet{sing_warm_2024} and \citet{welbanks_high_2024} suggest a core mass greater than 10 Earth masses, implying a need for an additional energy source in the deep atmosphere, which can keep its radius inflated. Possible sources for this internal heating include equilibrium tide dissipation \citep{welbanks_high_2024}, obliquity tide dissipation \citep{Leconte10, Millholland19, millholland_tidal_2020, Millholland20}, and Ohmic dissipation \citep{batygin_tides_2025, dyrek_so2_2024}.

As discussed in previous studies \citep{piaulet_wasp-107bs_2021, yu_are_2024}, WASP-107~b is a key target for understanding the role of inflation in the observed diversity of exo-Neptune densities. The dissipation of stellar tidal disturbances within a planet can provide a significant energy source, inflating the planet as it adjusts to a new thermal equilibrium and slowing its contraction. While \citet{Bodenheimer2001} invoked tidal inflation and non-zero eccentricity to explain the large radius of HD~209458~b, subsequent secondary eclipse observations confirmed its orbit was circular \citep{Deming2005}. 
In contrast, the measured non-zero eccentricity ($e_b \simeq 0.09$) for WASP-107~b raises the possibility that ongoing eccentricity tidal dissipation provides an internal heat source that helps maintain its large radius.
We therefore discuss tidal heating as a potential mechanism for the radius inflation of WASP-107~b \citep{piaulet_wasp-107bs_2021}. 
Unlike the discussion in \citet{batygin_tides_2025}, our analysis here focuses on the tidal energy injected into, and the heat flux released from, the core of WASP-107~b. The core lies at the bottom of the planetary atmosphere and heat flux released there plays a more significant role in driving the planet's radius expansion \citep{Guillot10, Komacek17}.
For planetary tides, the eccentricity-driven dissipation rate in the core of WASP-107~b can be approximated \citep{Jackson2008, Leconte2010} as:
\begin{equation}
\dot{E}_{\rm tide} \simeq \frac{21}{2}\,\frac{k_2}{Q_p}\,\frac{G M_\star^2 R_{\rm p,core}^5 n\,e^2}{a^6}\,,
\label{eq:tidal}
\end{equation}
where $k_2$ is the planet core's Love number, $Q_p$ its tidal quality factor of the core of WASP-107~b, $n=2\pi/P$ the mean motion, and $a$ the semi-major axis. It is custom to combine $Q_p$ and $k_2$ into the so-called `reduced tidal quality factor', $Q' = 3Q_p/2k_2$.

To assess the significance of this internal heating, we compare $\dot{E}_{\rm tide}$ with the estimated internal heat flux from the core of WASP-107~b,
\begin{equation}
P_{\rm core, rad}\simeq 4\pi R_{\rm p, core}^2 \sigma_{\rm sb} T^4_{\rm core} \simeq 4.7\times10^{18}\ \mathrm{W},\,,
\label{eq:tidal}
\end{equation}
where we adopt $T_{\rm core} = 460$~K and $R_{\rm p, core}= 1.9~R_\oplus$ for a $12M_\oplus$ heavy-element core \citep{sing_warm_2024,batygin_tides_2025}.
Adopting representative values of the core of WASP-107~b (orbital period $P\approx5.7$\,d, $e=0.09$, $M_\star\sim0.7\,M_\odot$, and $R_{\rm p, core}\sim1.9\,R_{\oplus}$) and equating $\dot{E}_{\rm tide}$ to $P_{\rm core, rad}$, we find the modified tidal quality factor $Q' \simeq 50$ for the core of WASP-107~b. 
This is consistent with the values for terrestrial planets in the Solar system, and typical core material for ice giants \citep{Goldreich66, welbanks_high_2024}.

Furthermore, as suggested by \citet{Bodenheimer2001}, radius inflation can intensify star-planet tidal interactions, which can effectively reduce $Q_p$. Thus, our preliminary estimates indicate that tidal energy could contribute significantly to the internal heating required by WASP-107~b, representing a substantial fraction of its internal radiation budget. This tidal dissipation deposits energy directly into the planetary interior, potentially providing the energy source needed to substantially slow the planet's Kelvin–Helmholtz contraction and maintain its inflated radius \citep[e.g.,][]{Bodenheimer2001, Guillot2002, Fortney2010}.
Thus, for WASP-107\,b, tidal heating at the estimated level could plausibly contribute to maintaining its inflated radius, especially if the dissipation is long-lived or the planet's thermal relaxation time is comparable to the tidal circularization timescale.

Our order-of-magnitude estimate also contains important caveats. First, $\dot{E}_{\rm tide}$ is linearly sensitive to the poorly constrained factor $Q'$; if $Q'$ is significantly larger than $10^{2}$, the tidal power becomes negligible. Second, this calculation does not consider other effects, such as high atmospheric opacities, Ohmic dissipation, obliquity tide, or reduced cooling from heavy-element enrichment and efficient day–night heat redistribution. Third, the interior location of the tidal energy deposition remains uncertain. Future work that couples interior evolution models with the measured orbital parameters, exploring a realistic range of $k_2$ and $Q_p$, will be needed to determine whether tidal heating can fully account for WASP-107~b's super-puffed radius or only provides a partial contribution.

\subsection{A possibly young exoplanet?}

There remain several outstanding mysteries about the WASP-107 system. One key issue is the apparent age discrepancy: different methods yield significantly different stellar ages. This inconsistency has been attributed to the strong magnetic activity of the host star, which may inhibit convective energy transport and lead to an inflated stellar radius, thereby biasing age estimates based on standard stellar models \citep[e.g.,][]{Feiden2016, mocnik_starspots_2017,Jackson2018}. The magnetic inhibition of convection is known to alter stellar structure, particularly in late-type stars, and could therefore explain both the larger-than-expected radius and the high chromospheric activity observed in WASP-107.

An alternative explanation worth exploring for the observed characteristics of the system, including the large planetary radii, the non-zero orbital eccentricity, and the host star’s high activity level, is that WASP-107 may be intrinsically young. \citet{Sun2022} proposed that WASP-107 is likely associated with a young stellar moving group, with a membership probability of about $70\%$. Although current evidence for this association remains tentative, the youth scenario provides a plausible framework for interpreting several of the system’s key properties, including the planet’s inflated radius and the persistence of a measurable eccentricity. The combination of residual tidal heating \citep{Bodenheimer2001, Guillot2002} and a still-contracting planet, magnetically active host star could thus naturally account for the system’s observed anomalies.

\section{Conclusion\label{sec:conclusion}}

We present a comprehensive joint analysis combining the JWST primary transit, JWST secondary eclipse, and Keck/HIRES radial velocity data for WASP-107~b. Our new joint fitting confirms that WASP-107~b possesses a small but significant non-zero orbital eccentricity of $e_b = 0.09 \pm 0.02$. 
Our new eccentricity measurement for WASP-107~b provides tentative evidence that the system may be in the final stage of high-eccentricity migration, supporting the formation scenario proposed by \citet{yu_are_2024}.
While further verification via more sophisticated modeling is needed, preliminary estimates indicate that eccentricity tidal dissipation could serve as one of the primary energy sources sustaining the exceptionally large inflated radius of WASP-107~b. 
Future JWST secondary-eclipse observations of WASP-107~b would be highly valuable for further refining its orbital eccentricity. 
Our results also provide timing constraints that can help optimize the scheduling of such future JWST observations.

\section{Acknowledgments}
We thank the anonymous referee for valuable suggestions that improved our manuscript, particularly regarding the tests of the secondary-eclipse detection significance. 
We acknowledge the financial support from National Key R\&D Program of China (2024YFA1611801), NSFC grant 12073092, 12103098,  the science research grants from the China Manned Space Project (No. CMS-CSST-2025-A26), and the Earth 2.0 research funding from SHAO. 

\section{Data Availability\label{sec:data availability}}
All the JWST, HST, and TESS raw data used in this paper can be found in MAST: \dataset[10.17909/xcgb-8407]{http://dx.doi.org/10.17909/xcgb-8407}.
The configuration files, analysis notebooks, light curve, transit timing variation (TTV) and RV data, results, and source code associated with this work will be available on \href{https://github.com/bluecloud135/Eccentricity_WASP-107b_paper.git}{GitHub}\footnote{\href{https://github.com/bluecloud135/Eccentricity_WASP-107b_paper.git}{https://github.com/bluecloud135/Eccentricity\_WASP-107b\_paper.git}}.
The software packages utilized in this work include: \texttt{Eureka!~v0.10} \citep{bell_eureka_2022}, \texttt{exoTEDRF~v1.4.0} \citep{feinstein_early_2023,radica_awesome_2023}, \texttt{dynesty} \citep{speagle_dynesty_2020,koposov_joshspeagledynesty_2022},  \texttt{emcee} \citep{foreman-mackey_emcee_2013}, \texttt{Iraclis} \citep{Tsiaras2016, Tsiaras2016_55, Tsiaras2018}, \texttt{astropy} \citep{astropy_collaboration_astropy_2022}, \texttt{pandas} \citep{team_pandas-devpandas_2025}, \texttt{numpy}/\texttt{scipy} \citep{van_der_walt_numpy_2011}, and \texttt{matplotlib} \citep{hunter_matplotlib_2007}.

\clearpage
\bibliographystyle{aasjournal}
\bibliography{references}

\clearpage
\begin{appendix}
\setcounter{figure}{0}
\renewcommand{\thefigure}{A\arabic{figure}}
\setcounter{table}{0}
\renewcommand{\thetable}{A\arabic{table}}

\section{Joint Fitting including CORALIE RV Data\label{app:coralie}}

\begin{figure*}[ht]
    \centering
    \includegraphics[width=0.99\textwidth]{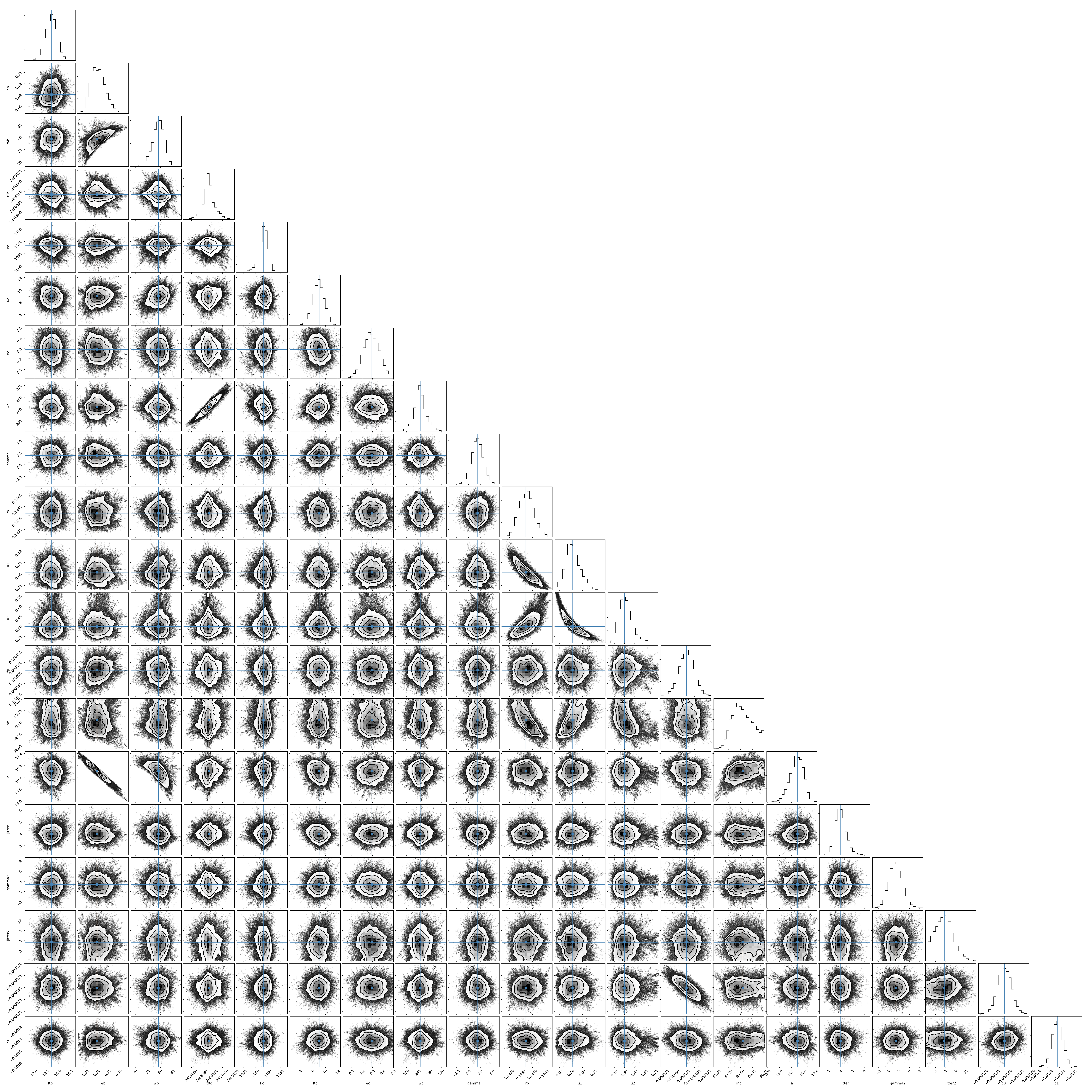}
    \caption{Corner plot for the mcmc fitting using both of the HIRES and CORALIE RV data. The limb-darkening coefficients $u_1$ and $u_2$ refer to the parameterization of \citet{kipping_efficient_2013}.}
    \label{fig:CORALIE_full_corner}
\end{figure*}

The CORALIE RV data from \citet{anderson_discoveries_2017} have relatively larger measurement errors and larger jitters comparing to the HIRES RV data, as shown in \citep{piaulet_wasp-107bs_2021}. Thus in our paper, we decided to use the HIRES RV data during our joint fitting study in Section 4. We have conducted additional joint fitting after incorporating the 31 CORALIE RV data points. The fitting results are shown in Table~\ref{tab:CORALIE} and Figure~\ref{fig:CORALIE_full_corner}. We do not find significant improvements upon the parameters of WASP-107~b comparing the results shown in Section~\ref{sec:ecc-measurement}. 
Most of the parameters are within 1-$\sigma$ to the best-fit results of \citet{piaulet_wasp-107bs_2021}.

\begin{deluxetable*}{lcc}[!h]
\tablecaption{MCMC joint-fit results for WASP-107 including additional CORALIE radial-velocity data. \label{tab:CORALIE}}
\tablewidth{0pt}
\tablehead{
\colhead{Parameter} & \colhead{Value} & \colhead{Error $(+\sigma\ / -\sigma)$ }
}
\startdata
$P_b$ (day) & $\equiv 5.7214876$ & $\pm0.0000001$ \\
$T_{0,b}$ (BJD-2450000) & $\equiv 9958.74727$ & $\pm3\times10^{-5}$ \\
$a/R_\star$  &  $16.5$  & $\pm0.04$ \\
$K_b$ (m s$^{-1}$) & 14.2 & $\pm0.8$ \\
$e_b$ & 0.09 & $\pm0.02$  \\
$e_b$ & 0.04  & 99.7\% lower limit \\
$e_b$ & 0.15  & 99.7\% upper limit \\
$\omega_b$ (deg) & 79.2 & +2.3 / -2.5 \\
$i_b$ (deg) & 89.53 & +0.27 / -0.20 \\
$M_b$ ($M_{\rm Jup}$) & 0.097 &$\pm0.005$ \\
$R_p/R_*$ & 0.1437 & $\pm0.0003$ \\
$F_p/F_*$ ($10^{-5}$) & 8.2 & $\pm1.8$ \\
$T_{0,c}$ (BJD-2450000) & 8932 & +45 / -37 \\
$P_c$ (days) & 1084 & +15 / -19 \\
$K_c$ (m s$^{-1}$) & 8.9 & $\pm0.9$ \\
$e_c$ & 0.29 &  $\pm0.08$ \\
$\omega_c$ (deg) & 246 & +27 / -30 \\
$M_c \sin i$ ($M_{\rm Jup}$) & 0.34 & $\pm0.04$ \\
$u_1$ & 0.15 & $\pm0.03$ \\
$u_2$ & 0.10 & $\pm0.06$ \\
$\gamma_H$ (m s$^{-1}$) & 1.3 & $\pm0.7$ \\
$\gamma_C$ (m s$^{-1}$) & 2.0 & $\pm2.0$ \\
$\sigma_H$ (m s$^{-1}$) & 4.0 & +0.5 / -0.4 \\
$\sigma_C$ (m s$^{-1}$) & 5.3 & +2.6 / -2.9 \\
$c_0$ (ppm) & -51 & $\pm13$  \\
$c_1$ (ppm/day) & -1.43E-3 & $\pm9$E-5 \\
\enddata
\tablecomments{
    Parameter definitions: 
    $K_b$/$K_c$ = Radial velocity semi-amplitude of component b/c; 
    $e_b$/$e_c$ = Orbital eccentricity of component b/c; 
    $\omega_b$/$\omega_c$ = Argument of periastron of component b/c; 
    $T_{0,b}$/$T_{0,c}$ = Transit/mid-transit time of component b/c (relative to BJD-2450000); 
    $P_c$ = Orbital period of component c; 
    $\gamma_{H, \; C}$ = Systemic radial velocity for HIRES and CORALIE data sets; $\sigma_{H, \; C}$ are the HIRES and CORALIE RV jitters term;
    $R_p/R_*$ = Planet-to-star radius ratio; 
    $u_1$/$u_2$ = Quadratic limb-darkening coefficients; 
    $F_p/F_*$ = Planet-to-star flux ratio; 
    $i$ = Orbital inclination; $c_0$ and $c_1$ are the linear systematic parameters for JWST occultation data.
}
\end{deluxetable*}

\section{Additional Table and Plots\label{app:more_table_figure}}

\begin{deluxetable*}{ccccccc}[htbp]
\tablecaption{Observation Proposal Information Used in This Work \label{tab:data proposal}}
\tabletypesize{\scriptsize} 
\tablewidth{0pt}
\tablehead{
\colhead{Planet} & \colhead{Mission}& \colhead{Instrument} & \colhead{Proposal ID} & \colhead{Observation} & \colhead{Proposal PI} & \colhead{Type}
}
\startdata
WASP-107~b & JWST & NIRCam F322W2 & 1185 & 8 & Thomas Greene & transit \\
WASP-107~b & JWST & NIRCam F444W & 1185 & 9 & Thomas Greene & transit \\
WASP-107~b & JWST & MIRI LRS & 1280 & 1  & Pierre-Olivier Lagage & transit \\
WASP-107~b & JWST & NIRISS SOSS\tablenotemark{a} & 1201 & 8  & David Lafreniere & transit \\
WASP-107~b & JWST & NIRSpec BOTS\tablenotemark{a} & 1224 & 3 & Stephan Birkmann & transit \\
WASP-107~b & JWST & NIRSpec BOTS\tablenotemark{a} & 1201 & 502  & David Lafreniere & eclipse \\
WASP-107~b & HST & WFC3 G141 & 14915 & - & Laura Kreidberg & transit \\
WASP-107~b & HST & WFC3 G102 & 14916 & - & Jessica Spake & transit \\
WASP-107~b & TESS & Photometer & DDT085 & - & Zixin Zhang \& Bo Ma & transit \\
\enddata
\tablecomments{
\tablenotetext{a}{For higher precision, we only used NRS1 detector data for NIRSpec observation and order 1 data for NIRISS observation.}
}
\end{deluxetable*}

\begin{deluxetable*}{ccc}[htbp]
\tablecaption{System parameters of WASP-107~b from previous studies \label{tab:WASP-107b parameters}}
\tabletypesize{\scriptsize}
\tablewidth{0pt}
\tablehead{
\colhead{Parameter} & \colhead{Value} & \colhead{Source}
}
\startdata
$R_p/R_{\star}$ & 0.1446 $\pm$ 0.0002 & \cite{kokori_exoclock_2023} \\
$P$ (days) & 5.72148926 $\pm$ 8.5e-07 & \cite{kokori_exoclock_2023} \\
$a/R_{\star}$ & 17.96 $\pm$ 0.05 & \cite{kokori_exoclock_2023} \\
$i$ ($^\circ$) & 89.56 $\pm$ 0.08 & \cite{kokori_exoclock_2023} \\
$M_p$ ($M_J$) & 0.096 $\pm$ 0.005 & \cite{piaulet_wasp-107bs_2021} \\
$R_p$ ($R_J$) & 0.924 $\pm$ 0.022 & \cite{mocnik_starspots_2017} \\
$M_{\star}$ ($M_\odot$) & 0.683 $\pm$ 0.017 & \cite{piaulet_wasp-107bs_2021} \\
$R_{\star}$ ($R_\odot$) & 0.67 $\pm$ 0.02 & \cite{piaulet_wasp-107bs_2021} \\
\enddata
\end{deluxetable*}

\begin{deluxetable*}{ccc}[htbp]
\tablecaption{Priors for transit light curve analysis \label{tab:LC priors}}
\tabletypesize{\scriptsize}
\tablewidth{0pt}
\tablehead{
\colhead{Parameter} & \colhead{Prior\tablenotemark{a}} & \colhead{Type}
}
\startdata
$R_p/R_{\star}$ & $\mathcal{U}$(0, 0.3) & uniform \\
$t_{\mathrm{mid}}$ & $\mathcal{N}$($t_0$, 0.01)\tablenotemark{b} & normal  \\
$P$ (days) & 5.72148926 & fixed\tablenotemark{c} \\
$a/R_{\star}$ & $\mathcal{N}$(17.96, 1) & normal \\
$i$ ($^\circ$) & $\mathcal{N}$(89.56, 1) & normal \\
$e$ & 0 & fixed\tablenotemark{c} \\
$u_1$ & $\mathcal{U}$(0, 1) & uniform \\
$u_2$ & $\mathcal{U}$(0, 1) & uniform \\
\enddata
\tablecomments{
\tablenotetext{a}{Based on Table \ref{tab:WASP-107b parameters}}
\tablenotetext{b}{$t_0$ from preliminary fit}
\tablenotetext{c}{Period ($P$) and eccentricity ($e$) are fixed for consistency with convention and computational efficiency; tests with free parameters showed negligible impact on results.
}}
\end{deluxetable*}

\begin{figure*}[ht]
    \centering
    \includegraphics[width=0.99\textwidth]{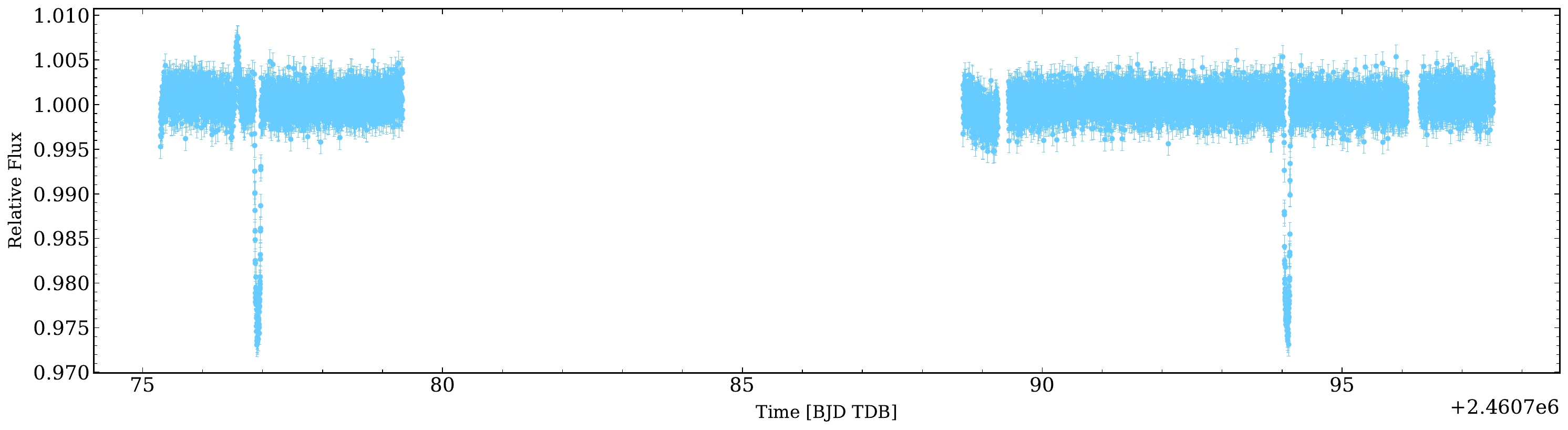}
    \caption{Full TESS light curve for WASP-107~b}
    \label{fig:WASP-107b full white light curve TESS}
\end{figure*}

\begin{figure*}[ht]
    \centering
    \includegraphics[width=0.49\textwidth]{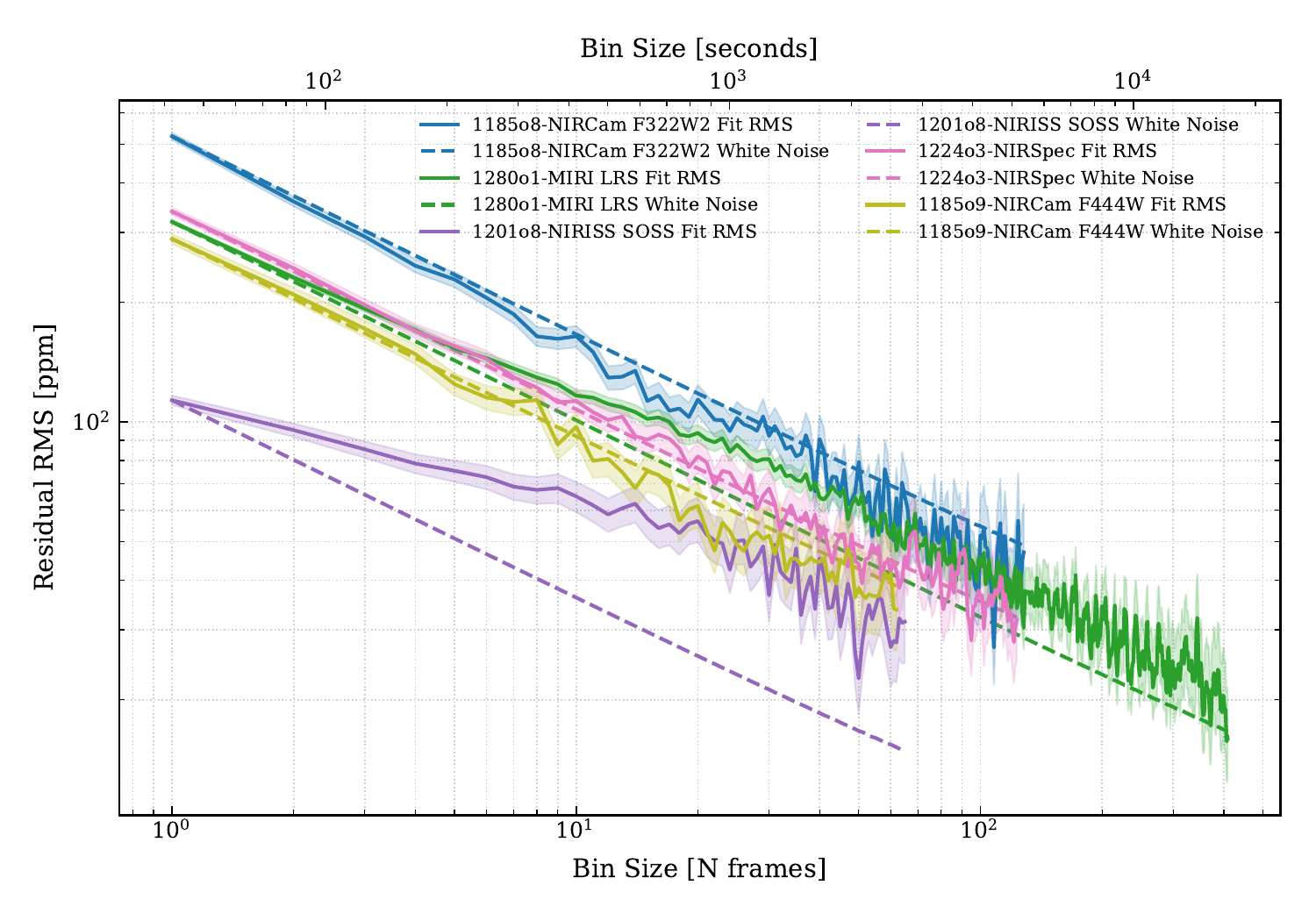}
    \includegraphics[width=0.49\textwidth]{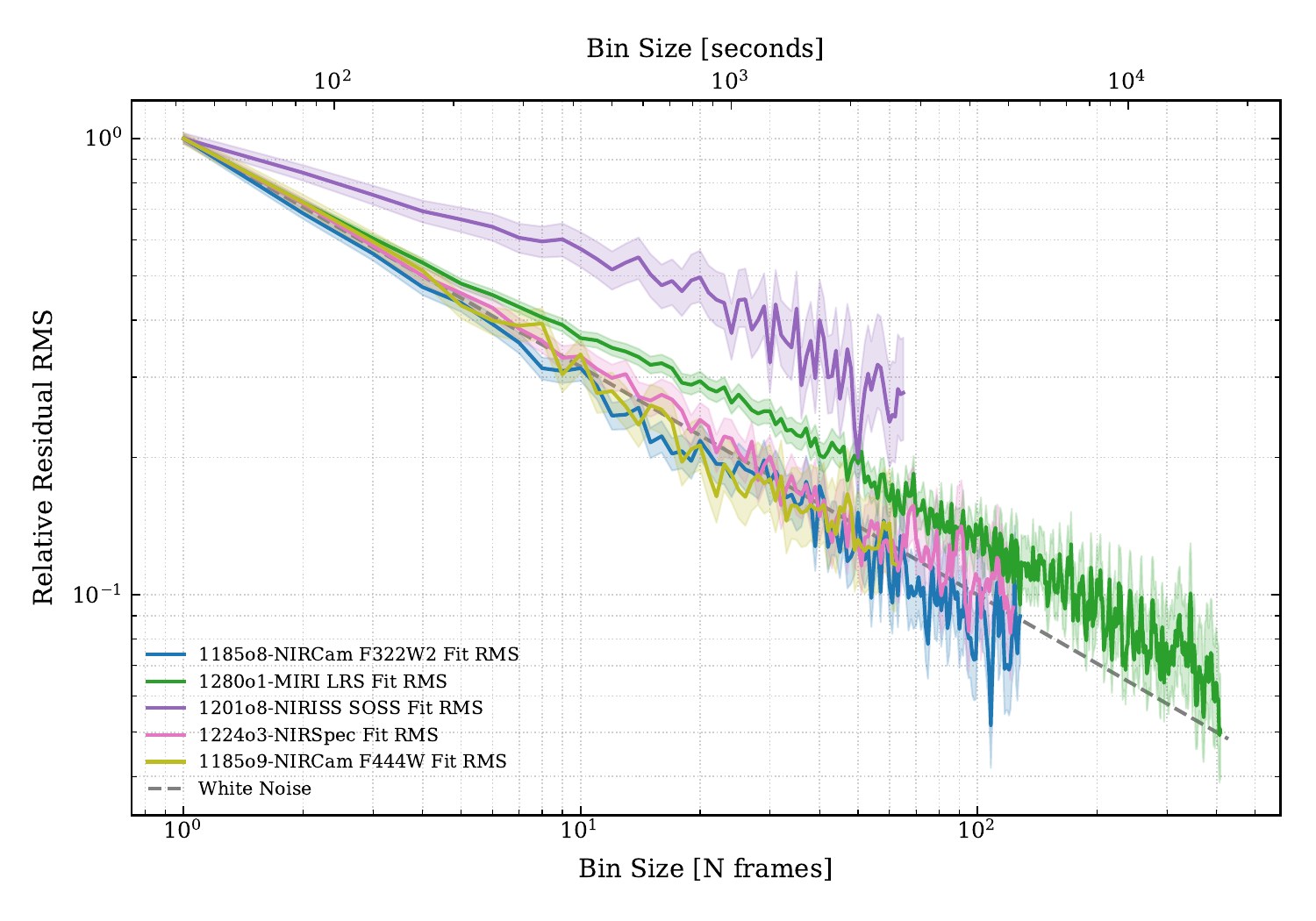}
    \caption{Time-averaging tests for JWST light curves. Left: Residual RMS versus bin size. Dashed lines show expected white noise. Right: Relative residual RMS versus bin number $N$. Gray dashed line indicates $1/\sqrt{N}$ white-noise expectation.}
    \label{fig:WASP-107b residual rms JWST}
\end{figure*}

\begin{figure*}[ht]
    \centering
    \includegraphics[width=0.49\textwidth]{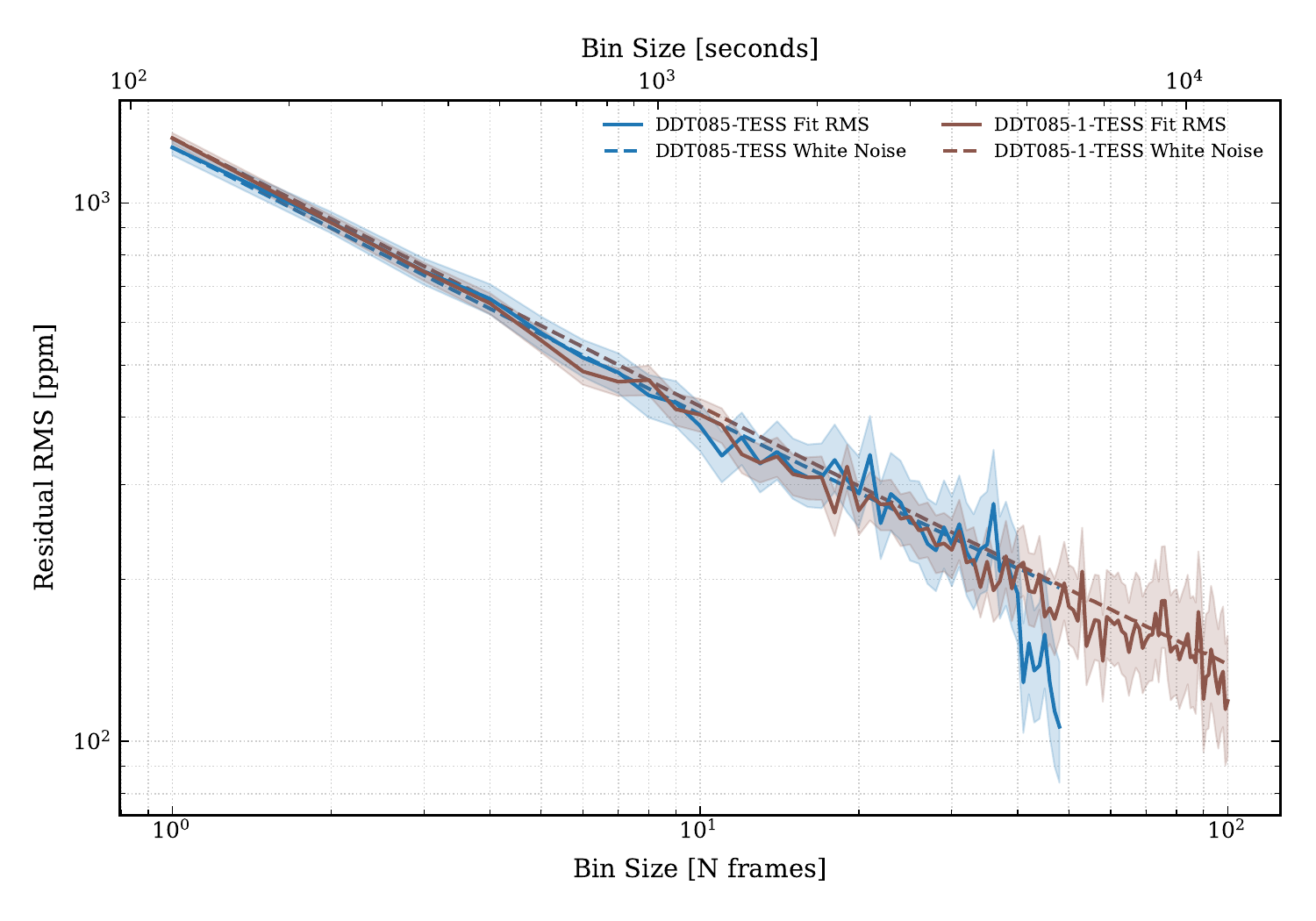}
    \includegraphics[width=0.49\textwidth]{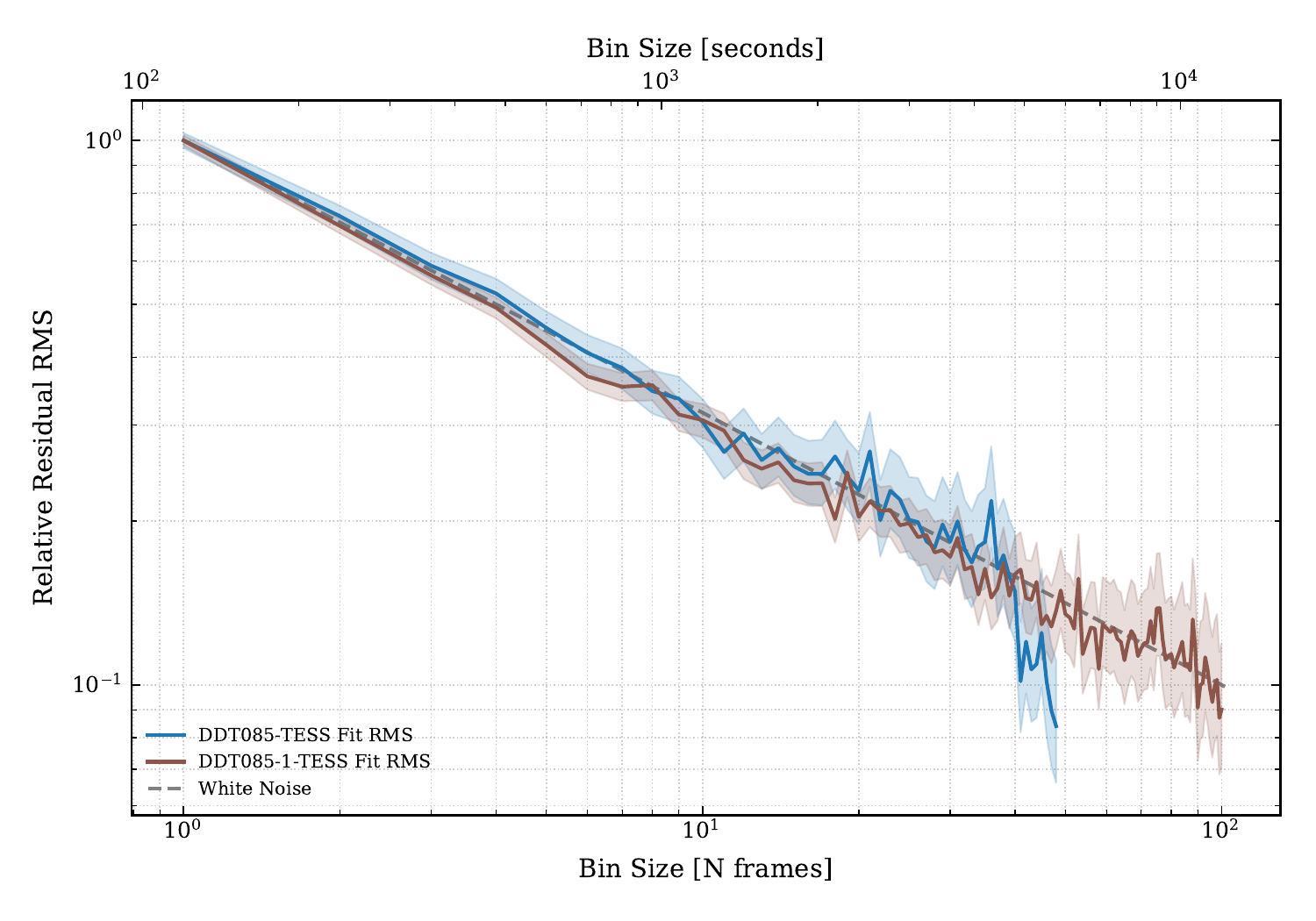}
    \caption{Time-averaging tests for TESS data (cf. Figure \ref{fig:WASP-107b residual rms JWST}).}
    \label{fig:WASP-107b residual rms TESS}
\end{figure*}

\begin{figure*}[ht]
    \centering
    \includegraphics[width=0.99\textwidth]{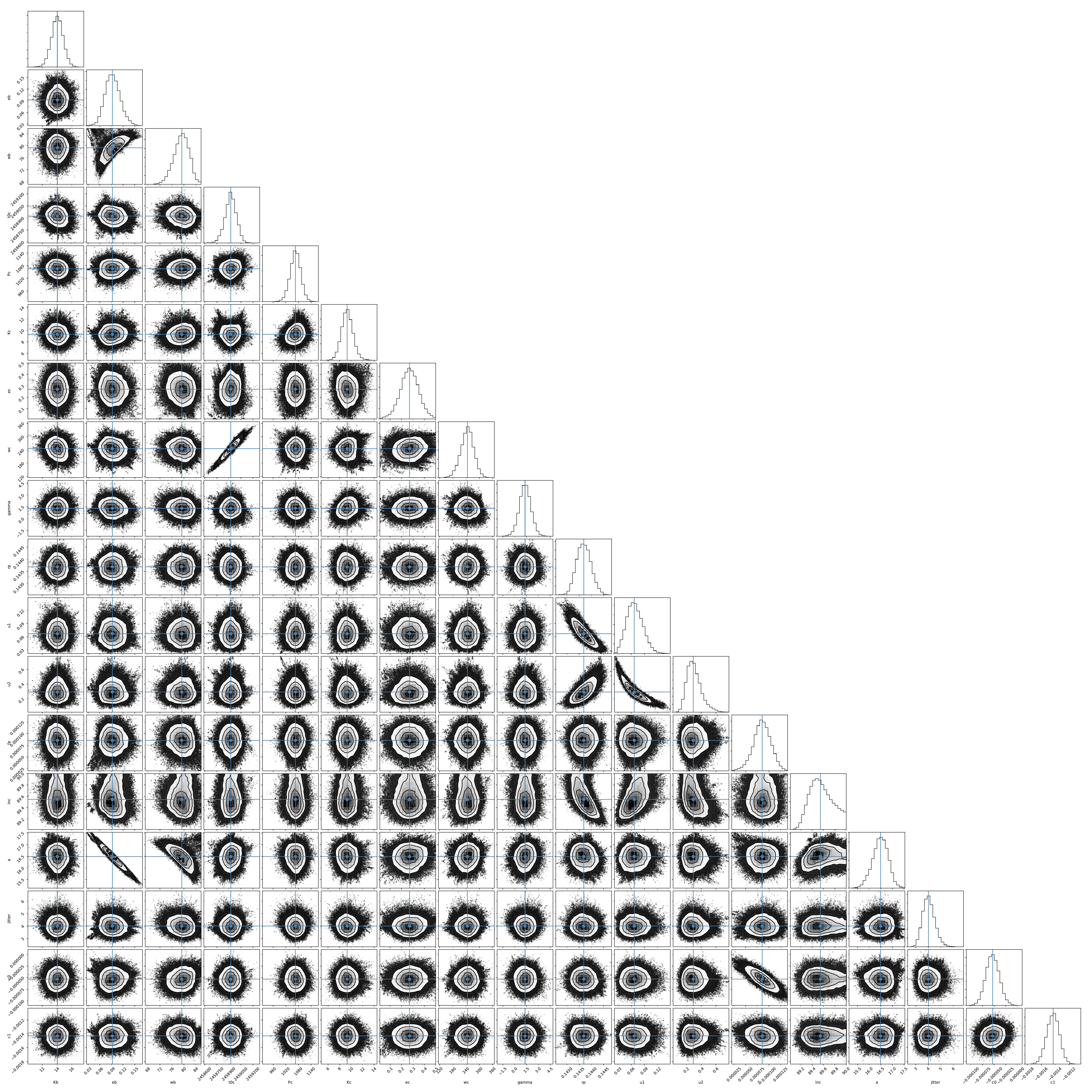}
    \caption{Corner plot for the mcmc fitting using only the HIRES RV data. The limb-darkening coefficients $u_1$ and $u_2$ refer to the parameterization of \citet{kipping_efficient_2013}.}
    \label{fig:HIRES_full_corner}
\end{figure*}

\end{appendix}

\end{document}